\documentstyle{article}\begin{document}
\title{Spectrum of a stochastic  diffusion in an expanding universe }
\author{ Z. Haba\\
Institute of Theoretical Physics, University of Wroclaw,\\ 50-204
Wroclaw, Plac Maxa Borna 9, Poland,
email:zhab@ift.uni.wroc.pl\\ Keywords:inflation,wave equation, power spectrum}\maketitle
\begin{abstract} We discuss the diffusion equation resulting from
a strong dissipation limit of the random wave equation arising in
the models of warm inflation. We show that the long wave power
spectrum of scalar perturbations in the model of  an exponential
expansion coincides with the spectrum of quantum fluctuations  on
this space-time.
\end{abstract}

\section{Introduction}
The standard consequence of the inflationary paradigm is an almost
scale invariant spectrum which is confirmed by WMAP observations
\cite{sto}\cite{ade}. The power spectrum is obtained by a
quantization of the quadratic fluctuations around the homogeneous
solution (derived first in \cite{muk}, see also the review
\cite{bas}). It has been pointed out \cite{bra1} that the almost
scale invariant spectrum may arise also in some versions of the
Pre-Big-Bang and Ekpyrotic cosmology \cite{khu1} \cite{khu2}
\cite{fin} as well as in the bounce models \cite{bra2}. The power
spectrum close to the scale invariant one has also been obtained
in warm inflation models \cite{hal}\cite{ram}. It is usually
believed that in the standard models of inflation the thermal
fluctuations are negligible \cite{yok}. However, such fluctuations
may be important in other cosmological theories \cite{bis}. In
most of these models the fluctuations satisfy a wave equation. In
this paper we study the stochastic wave equation in the form
derived from an interaction of scalar fields with an environment
\cite{ber1}, \cite{hab1}. We approximate  solutions of the
dissipative random wave equation by a solution of a diffusion
equation. In the case of a random diffusion the calculation of the
power spectrum is much simpler. Its dependence on the evolution
law can be seen in a more transparent way. We show that if the
evolution of the scale $a(t)$ is close to exponential and the
initial fluctuations are normalized to zero at $a=0$ then we
obtain the same power spectrum as in the quantum case when the
classical modes are normalized asymptotically to the Minkowski
plane waves \cite{bas}. This raises the question whether the
observed fluctuations \cite{sto} \cite{ade} are really of quantum
origin or perhaps they are thermal and classical. Beyond the
inflationary models the behaviour of quantum and thermal
fluctuations may be different. The plan of the paper is the
following. In sec.2 we discuss a dissipative stochastic wave
equation. In sec.3 we consider a diffusion approximation. In sec.4
we calculate the power spectrum under the assumption that the
evolution of the scale factor is almost exponential. In the
summary and outlook we discuss the differences between stochastic
and quantum fluctuations if the evolution is power-law. In the
Appendix we outline a method which allows to derive exact formulas
for the diffusion as well as wave equations if the evolution is
precisely exponential.

\section{Random wave equation}

In  a flat FLWR expanding metric
\begin{displaymath}
ds^{2}=dt^{2}-a^{2}d{\bf x}^{2},
\end{displaymath}
 we consider
the wave equation
\begin{equation}
\partial_{t}^{2}\phi_{\eta}-a^{-2}\triangle\phi_{\eta}+(3H+\gamma^{2})\partial_{t}\phi_{\eta}+V^{\prime}(\phi_{\eta})
+\frac{3r}{2}\gamma^{2}H\phi_{\eta}=\beta^{-\frac{1}{2}}\gamma
a^{-\frac{3}{2}}\eta,
 \end{equation}
where $\beta^{-1}$ is the temperature of the environment and
$H=a^{-1}\partial_{t}a$. The thermal noise is defined by the
covariance
\begin{equation}
\langle \eta_{s}({\bf x})\eta_{t}({\bf
y})\rangle=\delta(t-s)\delta({\bf x}-{\bf y}).
\end{equation}
This equation (with $r=1$) has been derived from an interaction of
the inflaton with an environment in \cite{ber1}\cite{hab1} and is
the basis of the warm inflation approach to cosmology \cite{ber2}
( then $r=0$ is assumed; $r$ may depend on the model of the
interaction of the inflaton with an environment).

We consider a linearized form of eq.(1) resulting from an
expansion around its homogeneous (space-independent) solution
\begin{equation}
\partial_{t}^{2}\phi_{ c}+(3H+\gamma^{2})\partial_{t}\phi_{c}+V^{\prime}(\phi_{c})
+\frac{3r}{2}\gamma^{2}H\phi_{c}=0,
 \end{equation}
We write  $\phi_{\eta}=\phi_{c}+\phi$. The initial conditions are
contained in $\phi_{c}$, so we assume zero as the intitial
condition for $\phi$. The linearization of eq.(1) expanded about
$\phi_{c}$ reads
\begin{equation}
\partial_{t}^{2}\phi-a^{-2}\triangle\phi+(3H+\gamma^{2})\partial_{t}\phi+V^{\prime\prime}(\phi_{c})\phi+\frac{3r}{2}\gamma^{2}H\phi=\beta^{-\frac{1}{2}}\gamma
a^{-\frac{3}{2}}\eta,
 \end{equation}
We can transform eq.(4) to another form . Let
\begin{equation}
\phi=a^{-\frac{3}{2}}\exp(-\frac{1}{2}\gamma^{2}t)\Phi
\end{equation}
Then

\begin{equation}
\partial_{t}^{2}\Phi-a^{-2}\triangle\Phi-\Omega^{2}\Phi=\gamma\exp(\frac{1}{2}\gamma^{2}t)\eta_{t}
 \end{equation}
where
\begin{displaymath}
\Omega^{2}=-V^{\prime\prime}+\frac{9}{4}H^{2}+\frac{3}{2}\partial_{t}H
+\frac{3r}{2}\gamma^{2}H+ \frac{1}{4}\gamma^{4}.
\end{displaymath}Hence, the wave equation with friction is
transformed into a wave equation with a  complex mass $\Omega$.
Note that large $3H+\gamma^{2}$ means large $\Omega$.

\section{Diffusion approximation}
In this section we show that the diffusion approximation to eq.(4)
, i.e., the omission of $\partial_{t}^{2}\phi$,  is equivalent to
the neglect of fast decaying modes (for  a large $\Omega$) in the
solution of eqs.(5)-(6). The diffusion approximation to eq.(4) in
the momentum space reads
\begin{equation}
(3H+\gamma^{2})\partial_{t}\phi+a^{-2}k^{2}\phi+V^{\prime\prime}(\phi_{c})\phi+\frac{3r}{2}\gamma^{2}H\phi=\beta^{-\frac{1}{2}}\gamma
a^{-\frac{3}{2}}\eta,
 \end{equation}
On the other hand we may express the solution of eq.(6)(momentum
space) by means of the Green function $G$
\begin{equation}
\Phi(t)=\beta^{-\frac{1}{2}}\gamma\int ds
G(t,s)\exp(\frac{1}{2}\gamma^{2}s)\eta_{s}ds\end{equation}where
the approximate Green function (for large slowly varying $\omega$)
is
\begin{equation}
G(t,s)=\omega(s)^{-\frac{1}{2}}\omega(t)^{-\frac{1}{2}}\sinh\Big(\int_{s}^{t}d\tau\omega(\tau)\Big)
\end{equation}with
\begin{equation}
\omega^{2}=\Omega^{2}-a^{-2}{\bf
k}^{2}=-V^{\prime\prime}(\phi_{c})-a^{-2}{\bf
k}^{2}+\frac{9}{4}H^{2}+\frac{3}{2}\partial_{t}H+\frac{1}{4}\gamma^{4}-\frac{3r}{2}\gamma^{2}H.
\end{equation}
Expanding $\omega$ in powers of $(3H+\gamma^{2})^{-1}$ we obtain
in the lowest order of the expansion
\begin{equation}\begin{array}{l}
\omega=\frac{3}{2}H+\frac{1}{2}\gamma^{2}-(3H+\gamma^{2})^{-1}(V^{\prime\prime}(\phi_{c})+a^{-2}{\bf
k}^{2}-\frac{3}{2}\partial_{t}H+\frac{3r}{2}\gamma^{2}H)\cr\equiv
\frac{3}{2}H+\frac{1}{2}\gamma^{2}+
(3H+\gamma^{2})^{-1}\frac{3}{2}\partial_{t}H -\mu,
\end{array}\end{equation} where
\begin{displaymath}\begin{array}{l}
\mu=(3H+\gamma^{2})^{-1}(V^{\prime\prime}(\phi_{c})+a^{-2}{\bf
k}^{2}+\frac{3r}{2}\gamma^{2}H).
\end{array}\end{displaymath}

We compare solutions of the wave equation (1) with solutions of
the diffusion equation (7).

The solution of the diffusion equation (7) is
\begin{equation}
\phi_{t}=\beta^{-\frac{1}{2}}\gamma \exp\Big(-\int_{s}^{t} d\tau
\mu(\tau)\Big)(3H(s)+\gamma^{2})^{-1}a(s)^{-\frac{3}{2}}\eta(s)ds
\end{equation}
We compare the solution (12) with (7)-(8). In the Green function
(9) we have
\begin{equation}\begin{array}{l}
\int_{s}^{t}d\tau\omega(\tau)=\frac{3}{2}\ln a(t)-\frac{3}{2}\ln
a(s)+\frac{1}{2}\gamma^{2}(t-s)-\frac{1}{2}\ln(3H(t)+\gamma^{2})\cr+\frac{1}{2}\ln(3H(s)+\gamma^{2})
 -\int_{s}^{t}d\tau \mu(\tau)
\end{array}\end{equation}
If in
\begin{displaymath}
\sinh(X)=\frac{1}{2}\exp(X)-\frac{1}{2}\exp(-X)
\end{displaymath}
we neglect the second term as quickly vanishing (for $X>0$) then a
simple comparison of eqs.(8)-(9) and (12)-(13) leads to the
conclusion that for large $3H+\gamma^{2}$ the solutions of the
wave equation and the diffusion equation (with zero initial
conditions) coincide.

\section{The power spectrum}
 The power spectrum $\rho$ of fluctuations $\phi$ is defined by the Fourier transform

\begin{equation} \langle \phi_{t}({\bf x})\phi_{t}({\bf y})\rangle
=\int d{\bf k}\rho_{t}({\bf k})\exp(i{\bf k}({\bf x}-{\bf y}))
\end{equation}or in Fourier  transform
\begin{equation}
\langle \phi_{t}({\bf k})\phi_{t}({\bf
k}^{\prime})\rangle=(2\pi)^{3} \delta({\bf k}+{\bf
k}^{\prime})\rho_{t}({\bf k}).\end{equation} The spectral index
$2\sigma$ is defined by the low $k=\vert {\bf k}\vert$ behaviour
$\rho_{t}({\bf k})\simeq k^{-2\sigma}$.
 The solution $\phi_{\eta}=\phi_{c}+\phi$ of eq.(1) with a given initial condition is a sum of the solution $\phi_{c}$
 of the
 homogeneous equation (3) with this initial condition and $\phi$ with 0 as an initial condition at $t_{0}$
 . Explicitly

\begin{equation}\begin{array}{l}
\phi_{t}=\beta^{-\frac{1}{2}}\gamma\int_{t_{0}}^{t}\exp\Big(-\int_{s}^{t}(3H+\gamma^{2})^{-1}(k^{2}a^{-2}+v(s^{\prime})
+\frac{3r}{2}\gamma^{2}H)ds^{\prime}\Big)
\cr a(s)^{-\frac{3}{2}}(3H+\gamma^{2})^{-1}\eta ds.
\end{array}\end{equation} where
\begin{equation}
v(s)=V^{\prime\prime}(\phi_{c}(s))
\end{equation}

Hence, \begin{equation}\begin{array}{l} \rho_{t}({\bf k})
=\beta^{-1}
\gamma^{2}\int_{t_{0}}^{t}\exp\Big(-2\int_{s}^{t}(3H+\gamma^{2})^{-1}(k^{2}a^{-2}
+v(s^{\prime})+\frac{3r}{2}\gamma^{2}H)ds^{\prime}\Big) \cr
a(s)^{-3}(3H+\gamma^{2})^{-2}ds. \end{array}\end{equation}  If we
introduce the e-fold time
\begin{equation}
d\nu=Hdt,
\end{equation}
then\begin{equation}\begin{array}{l} \rho_{t}({\bf k}) =\beta^{-1}
\gamma^{2}\int_{\nu(t_{0})}^{\nu(t)}(3H+\gamma^{2})^{-2}\exp\Big(-2\int_{\tau}^{\nu(t)}(3H^{2})^{-1}(1+\Gamma)^{-1}\Big(k^{2}
\exp(-2\tau^{\prime}) \cr
+v(\tau^{\prime})+\frac{3r}{2}\gamma^{2}H\Big)d\tau^{\prime}\Big)H^{-1}
\exp(-3\tau)d\tau,\end{array}
\end{equation} where

\begin{equation}
\Gamma=(3H)^{-1}\gamma^{2}.
\end{equation} We introduce
the variable
\begin{equation}
u=\exp(-2\tau)
\end{equation}
and assume that $H(\nu)\simeq const$ and $ m(\nu)^{2}H(\nu)^{-2}
\simeq const$ then
\begin{equation}\begin{array}{l}
\rho_{t}({\bf k})=
(\frac{1}{2H})^{2}\beta^{-1}\gamma^{2}\exp\Big((3H^{2})^{-1}(1+\Gamma)^{-1}k^{2}\exp(-2\nu)\Big)
\cr\int_{u(t_{0})}^{u(t)}\exp\Big(-(3H^{2})^{-1}(1+\Gamma)^{-1}k^{2}u-2q\nu\Big)
u^{\frac{1}{2}-q}du, \end{array}\end{equation}where
\begin{equation}
q=(\delta+\frac{3r}{2}\Gamma)(1+\Gamma)^{-1}\end{equation} and
\begin{equation}
\delta=v(3H^{2})^{-1}.
\end{equation}
 The result of
integration in eq.(23)can be expressed by the incomplete $\Gamma$
function
\begin{equation}\begin{array}{l}\rho_{t}({\bf k})
=(\frac{1}{2H})^{2}(3H+\gamma^{2})^{-2}\exp(-2q\nu)
\gamma^{2}\beta^{-1}\exp\Big((3H^{2})^{-1}(1+\Gamma)^{-1}k^{2}\exp(-2\nu)\Big)\cr
\Big(\Big((3H^{2})^{-1}k^{2}(1+\Gamma)^{-1}\Big)^{-\sigma}\Gamma(\sigma,(3H^{2})^{-1}(1+\Gamma)^{-1}k^{2}\exp(-2\nu))
\cr-\Big((3H^{2})^{-1}(1+\Gamma)^{-1}k^{2}\Big)^{-\sigma}\Gamma(\sigma,(3H^{2})^{-1}
(1+\Gamma)^{-1}k^{2}\exp(-2\nu_{0}))\Big),\end{array}\end{equation}
where
\begin{equation}
\sigma=\frac{3}{2}-(\delta+\frac{3r}{2}\Gamma)(1+\Gamma)^{-1}.
\end{equation}
We have for $ x<<1$
\begin{equation}
\Gamma(\alpha,x)=\Gamma(\alpha)-x^{\alpha}\Big(n!(\alpha+n)\Big)^{-1},
\end{equation}
and for  $x>>1$
\begin{displaymath}
\Gamma(\alpha,x)=x^{\alpha -1}\exp(-x).
\end{displaymath}
If $\nu_{0}\rightarrow -\infty$ then the second term in eq.(26) is
vanishing and the first factor is dominating. In such a case for a
small $k$
\begin{equation}\begin{array}{l}\rho_{t}({\bf k})
\simeq k^{-2\sigma}.\end{array}\end{equation}
 More precisely the behaviour
(29) takes place if $k$ is small and
\begin{equation}
k(a(\nu_{0})H)^{-1}>>1.
\end{equation}If \begin{equation}
k(a(\nu_{0})H)^{-1}<<1,
\end{equation} (and $a(\nu_{0})\leq a(\nu)$)
then \begin{equation}\rho_{t}({\bf k}) \simeq const.
\end{equation}
The dependence of the power spectrum of the massive quantum fields
on the scale in de Sitter space is discussed in \cite{sta} and
this dependence in the case of the stochastic wave equation in
\cite{hal}. The spectral index of cosmological perturbations
$n_{S}$ is related to the fluctuations of $\phi$ as
\begin{equation}
n_{S}-1=2\sigma-3.
\end{equation}
At $\gamma=0$ the result (29) coincides with the power spectrum of
quantum fluctuations which are derived  by a calculation of
$\langle\phi^{2}\rangle$ in the Bunch-Davis vacuum
\cite{muk}\cite{tak}\cite{lyt}(sec.24.3) (normalized so that the
scalar modes behave as plane waves at large $k(aH)^{-1}$). It
follows from eq.(26) that the amplitude of thermal fluctuations
 is determined by $H$,
$\sigma$ (known from CMB measurements), $\beta$ and $\gamma$
(which this way would be fixed by $\rho_{t}({\bf k})$). On the
other hand the friction $\gamma$ is related (depending on the
model) to other measurable quantities as  ,e.g., the diffusion
constant \cite{hab1} \cite{hab2} or the density of radiation at
the end of inflation in the warm inflation scenario \cite{ber3}.
The theory shows that under the assumption of almost exponential
expansion both the quantum fluctuations and the thermal
fluctuations of the inflaton lead to
 almost the same spectral index ( this index is crucial
 for distinguishing various inflation models on the basis of observational data
 \cite{lin}\cite{bam}).  With the
 present observational sensitivity the contributions of the
 thermal and quantum  fluctuations of the inflaton may be indistinguishable
 if the expansion is almost exponential. We have calculated the power spectrum in an external
 expanding metric(assuming it is close to exponential). We did not discuss the fluctuations of the
gravity
 (see \cite{bas} ). With gravity fluctuations the power spectrum of thermal
 and quantum fluctuations may be different. The power spectrum of
 the stochastic wave equation has been calculated in \cite{hal}
 \cite{ram} who take into account
 gravitational fluctuations and apply some tools from
 thermodynamics to calculate the total energy-momentum and entropy
 of the inflaton-gravity system.
\section{Summary and outlook}The $\Lambda$CDM model with the inflationary scenario is the basic paradigm  of the
 contemporary cosmology. The measured power spectrum
is believed to result from an almost exponential expansion. The
results of this paper show that it may be difficult on the basis
of the present astronomical observations to distinguish the
contribution of quantum noise from the thermal noise if the
expansion is exponential .
 If $H$ is varying (as is the case in the alternative cosmological models \cite{bra1}\cite{khu1}
 \cite{khu2}\cite{fin})
  then
some more precise estimates in both the quantum calculations of
the spectrum as well as in the calculations in the warm inflation
models can be compared with observational data to distinguish the
contribution of quantum and thermal fluctuations. In the power-law
expansion $a\simeq t^{\alpha}$ (with $H=\frac{\alpha}{t}$) the
diffusion equation (7) reads
\begin{equation}\begin{array}{l}
\partial_{t}\phi=-\Big(t^{-2\alpha}(\frac{3\alpha}{t}+\gamma^{2})^{-1}k^{2}
+(V^{\prime\prime}+\frac{3r\alpha\gamma^{2}}{2t})(\frac{3\alpha}{t}+\gamma^{2})^{-1}\Big)\phi\cr
+\gamma(\frac{3\alpha}{t}+\gamma^{2})^{-1}t^{-\frac{3\alpha}{2}}\eta.
\end{array}\end{equation}
  The
exploration of the solution of  eq.(34) is more involved. However,
 direct conclusions are possible in some special cases. If in eq.(34) we neglect $V^{\prime\prime} $ and
$\gamma^{2}$in all factors $ (\frac{3\alpha}{t}+\gamma^{2})^{-1}$
(as well as the term $\frac{3r}{2}\gamma^{2}H$ ) then we get the
scale invariant power spectrum $\rho\simeq k^{-3}$. If the noise
$\eta$ on the rhs of eq.(1) is multiplied by
$\sqrt{3H+\gamma^{2}}$ (as in \cite{ber2}) and we neglect
$V^{\prime\prime}+\frac{3r}{2}H\gamma^{2}$ then the power index
will be $\sigma=\frac{3}{2}+\frac{1}{\alpha-1}$ exactly as in the
case of quantum fluctuations on a power-law expanding universe
\cite{abb}. The calculations in this paper depend only on the
expansion law. The results could well apply to any model of an
expanding matter, e.g., exploding stars.

 \section{Appendix:Exact formula for the exponential expansion} When $a(t)=\exp(Ht)$ then  the solutions of
the diffusion as well as the wave equation can be obtained with
precisely controlled approximations. Let us denote
\begin{equation}
M^{2}=V^{\prime\prime}+\frac{3r}{2}\gamma^{2}H.
\end{equation}and assume that $M^{2}$ can be approximated by a
constant. The solution of the linear diffusion equation with zero
initial condition is
\begin{equation}\begin{array}{l}
\phi_{t}=\beta^{-\frac{1}{2}}\gamma\int_{t_{0}}^{t}ds\frac{1}{3H+\gamma^{2}}\eta_{s}\exp(-\frac{3}{2}H(t-s))\exp\Big(-\frac{k^{2}}{3H+\gamma^{2}}
\cr(\exp(-2Hs)-\exp(-2Ht))-\frac{M^{2}}{3H^{2}+H\gamma^{2}}(t-s)
\Big)\end{array}
\end{equation}
Let
\begin{displaymath}
u(s)=\exp(-2Hs)
\end{displaymath}
\begin{equation}
R=3H^{2}(1+\frac{1}{3}\gamma^{2}H^{-1})=3H^{2}(1+\Gamma)
\end{equation}
Then\begin{equation} \rho_{t}({\bf
k})=\frac{\gamma^{2}}{2H\beta}(\frac{1}{3H+\gamma^{2}})^{2}
\exp(\frac{k^{2}\exp(-2Ht)}{R})
\int^{u(t)}_{u(t_{0})}\exp(-\frac{k^{2}u}{R}) u^{\sigma-1}du
\end{equation} where
\begin{equation}
\sigma=\frac{3}{2}-\frac{M^{2}}{R}
\end{equation}
We have in eq.(38) the same integral as  in
eq.(26)\begin{equation}
 \rho_{t}({\bf
k})=\frac{\gamma^{2}}{2H\beta}
(\frac{1}{3H+\gamma^{2}})^{2}\exp(\frac{k^{2}\exp(-2Ht)}{R})
(\frac{k^{2}}{R})^{-\sigma}
\Big(\Gamma(\sigma,\frac{k^{2}}{R}u(t))-\Gamma(\sigma,\frac{k^{2}}{R}u(t_{0}))\Big).\end{equation}
Hence, if $t_{0}=-\infty$ ,i.e.,$ a(t_{0})=0$, then
\begin{equation}
\rho_{t}({\bf k})\simeq k^{-2\sigma}
\end{equation}
In general,  if $k\rightarrow 0$ with $\frac{k}{a}=k\exp(-Ht_{0})$
bounded then still we have  $\rho_{t}({\bf k})\simeq k^{-2\nu}$
without referring to initial conditions ,i.e., we obtain the
behaviour discussed at eqs.(29)-(32). If $\gamma=0$ then
\begin{equation} \sigma =\frac{3}{2}-\delta
\end{equation}(with $\delta$ defined in eq.(25)).

This is exactly the index resulting from a quantization of  the
scalar field in an exponentially expanding universe
\cite{muk}\cite{tak}\cite{lyt}. Such a conclusion has also been
derived in warm inflation for $M=0$ in \cite{hal} .

We can estimate the solution of the stochastic wave equation in
the WKB approximation (9). Let
\begin{equation}
q=\frac{9}{4}H^{2}+\frac{1}{4}\gamma^{4}+\frac{3r}{2}\gamma^{2}H
-m^{2}.
\end{equation}
Now
\begin{equation}
\int_{s}^{t}\omega=-\frac{1}{2H}\int(q-k^{2}u)^{\frac{1}{2}}u^{-1}du
\end{equation}
or if we introduce
\begin{equation}
v=q-k^{2}u
\end{equation}
\begin{equation}\begin{array}{l}
\int_{s}^{t}
\omega=H^{-1}\sqrt{v(t)}-H^{-1}\sqrt{v(s)}-\frac{\sqrt{q}}{2H}\Big(2Ht-2Hs
+\ln\Big(2q-k^{2}u(t)+2\sqrt{q}\sqrt{v(t)}\Big)\cr
-\ln\Big(2q-k^{2}u(s)+2\sqrt{q}\sqrt{v(s)}\Big).
\end{array}
\end{equation}
Then, if we first expand $\int_{s}^{t} \omega$
 in  powers of $ \frac{1}{3H+\gamma^{2}}$ and subsequently calculate $\Phi$ according to eq.(9) then from the solution (9) of
 the wave equation we get the same result (42) for the power
 spectrum.

\end{document}